\documentclass[10pt,twocolumn]{revtex4-2}

\usepackage[letterpaper,
            top=0.65in,
            bottom=0.65in,
            left=0.65in,
            right=0.65in]{geometry}
\usepackage{amssymb,amsmath,graphicx,hyperref}
\newcommand{\beq}{\begin{equation}}
\newcommand{\eeq}{\end{equation}}
\newcommand{\bqa}{\begin{eqnarray}}
\newcommand{\eqa}{\end{eqnarray}}

\newcommand{\dg}{^\dagger}

\newcommand{\erf}[1]{Eq.~(\ref{#1})}

\newcommand{\erfand}[2]{Eqs.~(\ref{#1}) and (\ref{#2})}
\newcommand{\Erf}[1]{Equation~(\ref{#1})}
\newcommand{\smallfrac}[2]{\mbox{$\frac{#1}{#2}$}}

\newcommand{\ket}[1]{ |{#1} \rangle}
\newcommand{\Bra}[1]{\left\langle{#1}\right|}
\newcommand{\Ket}[1]{\left|{#1}\right\rangle}
\newcommand{\ip}[2]{\left\langle{#1}\right|\left.{#2}\right\rangle}
\newcommand{\sch}{Schr\"odinger}

\newcommand{\half}{\smallfrac{1}{2}}

\newcommand{\sq}[1]{\left[ {#1} \right]}
\newcommand{\cu}[1]{\left\{ {#1} \right\}}
\newcommand{\ro}[1]{\left( {#1} \right)}
\newcommand{\an}[1]{\left\langle{#1}\right\rangle}

\newcommand{\st}[1]{\left| {#1} \right|}

\newcommand{\kB}{k_{\rm B}}
\newcommand{\ti}{0}
\newcommand{\tip}{}
\newcommand{\tk}{k\delta t}
\newcommand{\dd}{{\rm d}}
\newcommand{\xfrac}[2]{{#1}/{#2}}

\begin{document}

\title{Entropy Flow of a Laser Beam}

\author{Howard M. Wiseman}
\email{h.wiseman@griffith.edu.au}
\affiliation{
Quantum and Advanced Technologies Research Institute, Griffith University, Yuggera Country, Brisbane, Queensland 4111, Australia}

\begin{abstract}A laser beam is often modelled by a pure coherent state. In fact its state is mixed, even if it has coherent-state photon-number statistics (Poissonian), because the phase must vary. We consider such an ideal laser beam, with phase diffusion rate $\ell$, equal to its (Lorentzian) spectral width. We show that the beam entropy is extensive, with an entropy flow of $\dot{S} = \kB\sqrt{\dot{N}\ell}$, where $\dot{N}$ is the number flow. We give an intuitive explanation for this remarkably simple result, and compare it to the entropy flow of a unidirectional thermal beam. \\ 
 \\
\noindent
\textbf{Published version: }
\href{https://doi.org/10.1364/OL.599036}{10.1364/OL.599036}
\end{abstract}

\maketitle

%%%%%%%%%%%%%%%%%%%%%%%%%%  body  %%%%%%%%%%%%%%%%%%%%%%%%%%
\section{Introduction}
\label{introduction}

Since the invention of the laser, its output has been recognized for, and found myriad applications due to, its superior coherence properties compared to other light sources~\cite{Wis16a}. For many purposes, a laser beam may be described as a perfectly coherent state $\ket{\alpha}$, with $\alpha$ constant in time, and in some quantum optics textbooks it is so described. However, more careful treatments show that even a laser with with no technical noise has quantum noise leading to phase diffusion, at some rate $\ell$, which is also equal to the FWHMH (full-width-at-half-maximum-height) spread in frequencies~\cite{WallsMilb94,MilonniEberly2010}. Since a laser beam is thus not in a pure state, this raises the question: what is its entropy? Surprisingly, to the author's knowledge, this has never been calculated. This paper rectifies that.

\section{Problem statement and solution}
%%\label{}
A free-space light-beam which is unidirectional, approximately monochromatic, polarized, and fully transverse-coherent is described by a 1-parameter family of bosonic lowering operators $\hat b(t)$~\cite{WallsMilb94,BSBW20}. Here, $t$ can be understood as the time since that infinitesimal beam-segment was emitted from the source, so that, at clock-time $\tau$, the field a distance $z$ from the laser is described by $\hat b(\tau -z/c)$. With commutation relations $[\hat b(t),\hat b(s)\dg] =\delta(t-s)$, the operator $\hat b\dg(t) \hat b(t)$ is the {\em photon flow} operator. We denote the mean photon flow, assumed time-independent, as $\dot{N}=\an{\hat b\dg(t)\hat b(t)}$ (rather than ${\cal N}$ as in Ref.~\cite{BSBW20}).

An ideal laser beam can be modelled as a constant-amplitude but wandering-phase coherent state. That is, an eigenstate of $\hat b(t)$, with eigenvalue $\sqrt{\dot{N}}e^{i\phi(t)}$, where $\phi(t)$ is a real stochastic process. Separating the beam into small segments of length $\delta t=T/K$, we can write this state over a duration $[\ti,\tip T)$ as
\beq \label{defrho}
\rho_{[\ti,\tip T)} = \lim_{K\to\infty}\mathbb{E}_\phi\left[\bigotimes_{k=0}^{K-1}  \Ket{re^{i\phi(\tk)}}\Bra{r e^{i\phi(\tk)}}\right],
\eeq
where $\mathbb{E}_\bullet$ denotes the ensemble average over $\bullet$, and the ket $\ket{re^{i\varphi}}$ is a single-mode coherent state of mean photon number $r^2 = \dot{N}\delta t$. We will consider pure phase diffusion,
\beq \label{phasewander}
\phi(t)=-\omega_0 t + \sqrt{\ell}W(t),
\eeq
where $\omega_0 \ (\gg \ell)$ is the centre-frequency and $W(t)$ is a Wiener process. This well describes quantum-limited phase noise, or other homogenous noise giving rise to a Lorentzian lineshape of FWHMH $\ell$~\cite{MilonniEberly2010}.  We also restrict to the usual regime of laser beams where the beam {\em coherence} $\mathfrak{C}$ --- the population of the maximally populated mode~\cite{BSBW20}  --- is extremely large. For an ideal laser, this evaluates to $\mathfrak{C}=4\dot{N}/\ell$~\cite{BSBW20}, and matches the intuitively defined {\em photon degeneracy} --- the number of photons per coherence volume~\cite{MilonniEberly2010} --- as introduced by Mandel in 1961~\cite{Mandel61}. 

We wish to calculate the entropy flow of the beam, 
\beq
\dot{S} = \lim_{T\to \infty} T^{-1} S[\rho_{[\ti,\tip T)}],
\eeq
which is independent of $\phi(\ti)$ and $T$. Because the von Neumann entropy, $S_1$, is too difficult to calculate, we use the R\'enyi 2-entropy, $S_2$. For the asymptotic-in-$T$ results, the beam effectively comprises many statistically independent identical segments of length $\approx 1/\ell$, so these entropies will differ at most by a constant factor~\cite{bobkov2018}. 
Thus, the quantity to be calculated is 
\beq \label{tobe1}
\dot{S}_2 = \lim_{T\to \infty} T^{-1} \cu{-\kB  \log {\rm Tr} \left[(\rho_{[\ti,\tip T)})^2\right]}
\eeq
As we prove below, \erf{tobe1}, the entropy flow of a laser beam of linewidth $\ell$ much less than its photon flow $\dot{N}$, equals
\beq \label{theanswer}
\dot{S}_2 = \kB\sqrt{\dot{N}\ell}
\eeq

%\beq \label{theanswer}
%{S}_2 \sim \sqrt{\dot{N}\ell}\, T%\kB
%\eeq

\section{Proof}

Being careful to use independent Wiener processes, say $W(t)$ and 
$V(t)$, for the two copies of $\rho$ in \erf{tobe1}, we have (where $r^2 = \dot{N}\delta t$ and $\delta t=T/K$),
\beq
{\rm Tr} \left[(\rho_{[\ti,\tip T)})^2\right] = \lim_{K\to\infty}\mathbb{E}_{W,V}
\sq{ \prod_{k=0}^{K-1} \st{\ip{re^{i\sqrt{\ell}W(\tk)}}{re^{i\sqrt{\ell}V(\tk)}} }^2}.  \label{evaluendum}
\eeq
Evaluating the inner product of these coherent states~\cite{WallsMilb94}, and defining a new Wiener process $X=(W-V)/\sqrt{2}$, the right-hand-side of \erf{evaluendum} evaluates to 
\begin{align}
& \lim_{K\to\infty}\mathbb{E}_{X} 
\sq{ \prod_{k=0}^{K-1} \exp\ro{-4r^2\sin^2\cu{\sqrt{\ell/2}\,X(\tk)}}}   \\
=& \lim_{K\to\infty}\mathbb{E}_{X} 
\sq{ \exp\ro{\sum_{k=0}^{K-1} -4r^2\sin^2\cu{\sqrt{\ell/2}\,X(\tk)}}}.
\end{align}
Now substituting in $r^2=\dot{N}\delta t$, and then taking the limit $K\to\infty$ so $\delta t\to\dd t$, gives 
\beq \label{that}
{\rm Tr} \left[(\rho_{[\ti,\tip T)})^2\right] = \mathbb{E}_{X} 
\sq{\exp\ro{-4\dot{N}\int_{\ti}^{\tip T} \sin^2\cu{\sqrt{\ell/2}\,X(t)}\dd t}}. 
\eeq 

This can be attacked by the Feynman--Kac technique~\cite{Kac1949}.  Let $X(0)=x$ %and $T=t$ 
and define the right-hand-side of \erf{that} as $u(x,T)$. Then it can be shown~\cite{borodin2002handbook,Lalley} that this function is the unique solution, satisfying $u(x,\ti)=1$, of %the ``heat diffusion equation'' in a periodic potential, 
\beq \label{Feynman-Kac}
 - \frac{\partial u}{\partial T} = {\cal L}u, \textrm{ where } {\cal L} := -\frac{1}{2}\frac{\partial^2}{\partial x^2} + 4\dot{N} \sin^2(\sqrt{\ell/2}\,x).
\eeq
 Now, the differential operator ${\cal L}$ is equivalent to the 1D \sch\ Hamiltonian for a $\sin^2$ potential, so clearly has a positive spectrum, meaning $u$ decays monotonically over time. The lowest eigenvalue, $\lambda_0>0$,  has a corresponding eigenfunction $u_0(x)$ that can be defined to be positive everywhere. 
  Thus, in the long-time limit, for all $x$, $u(x,T) \to c_0 u_0(x)e^{-\lambda_0 T}$, with $c_0>0$. So, whether integrating over a uniform distribution for $x$ [{\em i.e.} averaging in \erf{defrho} over all $\phi(0)$], or setting $x=0$ [{\em i.e.} assuming $\phi(0)$ to be fixed], \erf{tobe1} equals $\kB\lambda_0$. %The same is true for 
 
% u(x,0) = 1 = \sum_n c_n u_n(x). 
%   and the decay of the long-time solution $u(x,T)$ is determined by the smallest eigenvalue of ${\cal L}$. 
Now, to find $\lambda_0$, define a small parameter $\epsilon = (4\mathfrak{C})^{-1/4} = \half(\ell/\dot{N})^{1/4}$ 
 %(recall that $\mathfrak{C}=4\dot{N}/\ell$ and that the regime of interest is $\mathfrak{C} \gg 1$) 
 and change variables to $y= \sqrt{\ell/2}\,x/\epsilon$. Then 
 \beq
 {\cal L} =  2\sqrt{\dot{N}\ell} \left[ -\frac{1}{2}\frac{\partial^2 }{\partial y^2} + \frac{1}{2\epsilon^2} \sin^2(\epsilon y) \right].
\eeq
In the desired high coherence limit, $\mathfrak{C} \gg 1$, the term in square brackets can be approximated by the dimensionless harmonic oscillator Hamiltonian, with minimum eigenvalue of $1/2$. Thus $\lambda_0 = \sqrt{\dot{N}\ell}$, and we obtain \erf{theanswer}. 
 
\section{Discussion of \erf{theanswer}}

%There is much to say about \erf{theanswer}. 
First, while such an expression has never been given before, in hindsight it is quite reasonable. The only parameters that define the ideal laser beam are three rates: 
photon flow $\dot{N}$, linewidth $\ell$, and mean frequency $\omega_0$. The entropy of the mixture of coherent states is independent of $\omega_0$.  
%and this was, in fact, ignored in defining $\phi(t)$ below \erf{defrho}.
The entropy flow must go to zero as $\dot{N}\to 0$  (zero photon flow means a vacuum state, which is pure) and as $\ell\to 0$ (zero linewidth means a constant phase coherent state, which is pure). Therefore $\dot{S}_2= \kB  \sqrt{\dot{N}\ell}$ is the simplest expression we could hope for.

\begin{figure}[ht]
\centering
\fbox{\includegraphics[width=0.8\linewidth]{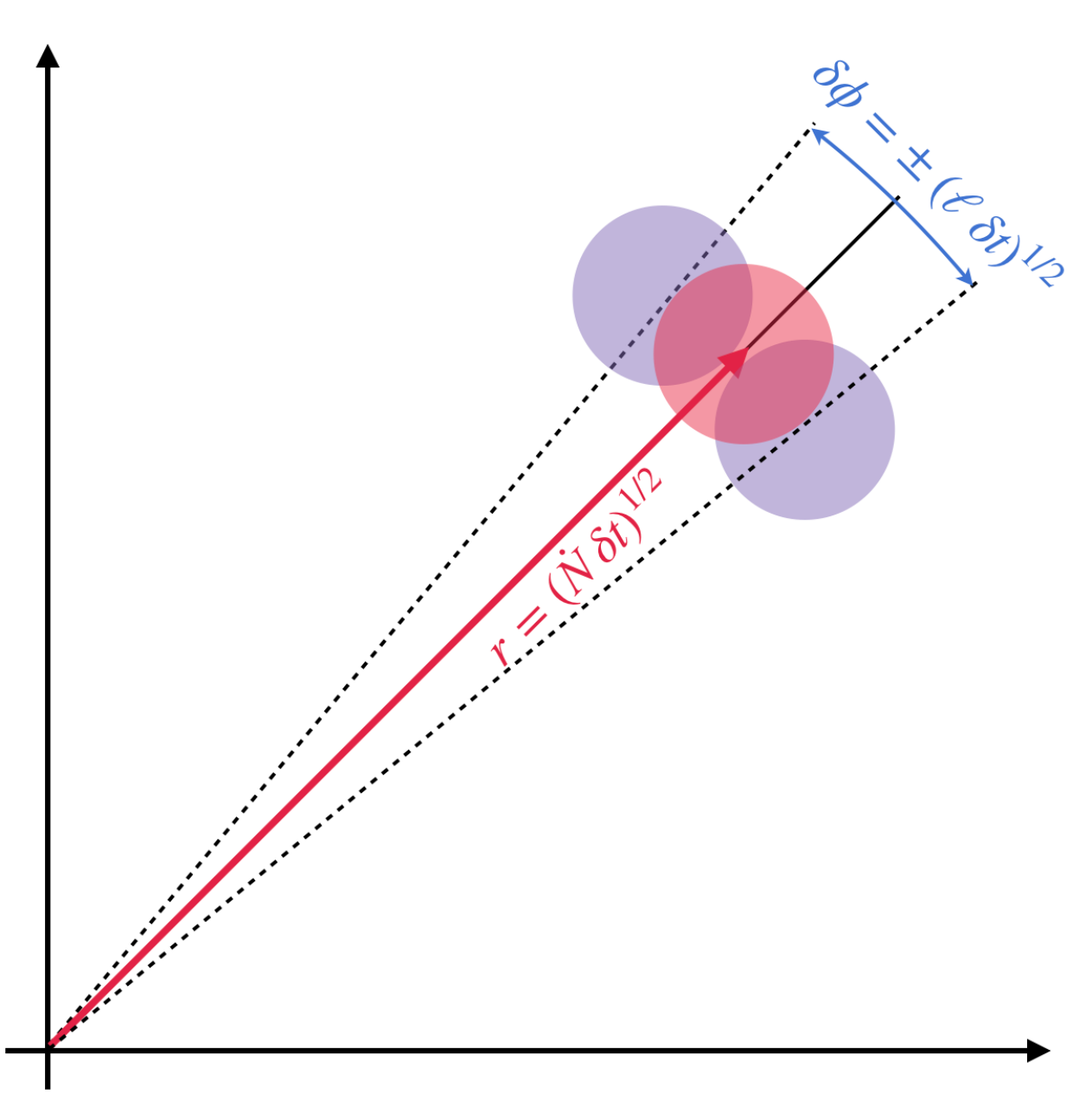}}
\caption{Phase-space diagram (arbitrary quadratures) showing phase diffusion (blue) from one $\delta t$-segment of laser beam (pink) to the next (mauve). The size of the colored circles indicate the quantum uncertainty of the coherent states. The approximate orthogonality of the two possible states in the second segment implies $\delta t \approx 1/\sqrt{\dot{N} \ell}$ and $\dot{S} \approx \kB/\delta t$.}
\label{fig:PhaseDiff}
\end{figure}

Second, we can understand \erf{theanswer} by thinking of the number of distinguishable states in the ensemble  (\ref{defrho}); see also Fig.~\ref{fig:PhaseDiff}. Each random walk of the phase through time gives a {\em different} state of the beam, but these are not necessarily distinguishable. They are if the coherent states $\ket{re^{i\phi(k\delta t)}}$, in some segment $k$, are close to orthogonal.  For this we must consider $K$ finite so that the modulus of the coherent amplitude $r=\sqrt{\dot{N}\delta t}$ is non-infinitesimal. Let $\delta t = 1/\sqrt{\dot{N}\ell}$, so that $r = (\dot{N}/\ell)^{1/4}$. Over this time, the random change in the phase is, roughly, $\pm \sqrt{\ell \delta t} = \pm (\dot{N}/\ell)^{-1/4}$. Since this is small, the complex coherent amplitude is displaced randomly (on the ring of radius $r$) by $\pm r\sqrt{\ell \delta t} = \pm 1$. This is just the displacement needed to make two coherent states roughly orthogonal. So, every time $\delta t$, a phase-path diverges into two distinguishable phase-paths. This increases the entropy by roughly $\kB\log(2)$, giving $\dot{S} \approx \kB/\delta t$, in agreement with \erf{theanswer}.

Third, for a ideal standard quantum-noise-limited laser~\cite{Wis99}, $\ell=\kappa/2\mu$ and $\dot{N}=\kappa\mu$, where $\mu$ is the mean number of {\em intracavity} photons and $\kappa$ is the bare cavity intensity-damping rate. This gives $\dot{S}_2/\kB = \kappa/\sqrt{2}$, clearly limited by quantum noise. 
Moreover, if the standard gain process were replaced by one with zero phase noise~\cite{Wis99}, one would have  $\ell=\kappa/4\mu$, yielding the even nicer expression $\dot{S}_2=\kappa/2$, equal to the bare cavity amplitude-damping rate.

Fourth, it is useful to rewrite \erf{theanswer} as
\beq \label{laserS}
\dot{S}_2 = \kB \sqrt{\xfrac{P}{\hbar}}\sqrt{\xfrac{\ell}{\omega_0}},
\eeq
where $P=\hbar \omega_0\dot{N}$ is the laser beam power, to compare with a thermal beam. A unidirectional polarised one-dimensional thermal beam at temperature $\Theta$ has a power~\cite{Wis16a}
\beq \label{thermalU}
P^\Theta = \dot{U}^\Theta =  \int_0^\infty \frac{\dd \omega}{2\pi}\,  \hbar \omega \, \bar{n}^\Theta_\omega  = \frac{\pi}{12} \frac{(\kB \Theta)^2}{\hbar}, 
\eeq
where $\bar{n}^\Theta_\omega = (e^{\hbar \omega/\kB\Theta}-1)^{-1}$. Its R\'enyi-2 entropy flow is 
\beq \label{thermalS}
\dot{S}_2^\Theta = \kB \int_0^\infty \frac{\dd \omega}{2\pi}\, 
\log(2\bar{n}^\Theta_\omega+1) = \kB \frac{\pi}{8} \frac{\kB \Theta}{\hbar},
\eeq
since a single-mode thermal state has ${\rm Tr}[\rho^2] = 1/(2\bar{n}+1)$. \Erf{thermalS} is just a factor of $3/4$ smaller than the von Neumann entropy flow, $\dot{S}^\Theta_1$, well known from 1+1--conformal field theory~\cite{Calabrese_2009}. 
%\dot{S}_2 = \kB \int_0^\infty \frac{\dd \omega}{2\pi}\, 
%\sq{(1+\bar{n}^\Theta_\omega)\log(1+\bar{n}^\Theta_\omega) - \bar{n}^\Theta_\omega \log \bar{n}^\Theta_\omega}.
Thus, combining \erfand{thermalU}{thermalS}, we have 
\beq
\dot{S}_2^\Theta = \kB \sqrt{\xfrac{P^\Theta}{\hbar}}\sqrt{\xfrac{3\pi}{16}}.
\eeq
Comparing to \erf{laserS}, we see that, for the same power, the laser beam has an entropy smaller by a factor of roughly $Q^{-1/2}$, where $Q=\omega_0/\ell$ is the Q-factor of the beam. Many commercial solid-state lasers have $Q \approx 10^{10}$ or better~\cite{Zhao2021}. 
%, meaning the laser:thermal beam-entropy ratio can easily be $10^{5}$. 
To give a numerical value, for $Q =10^{10}$ and $P = 750$mW we have $\dot{S}_2 \approx 10^{-12}$W/K or, perhaps more usefully, $\sqrt{\dot{N}\ell} \approx 10^{11}{\rm s}^{-1}$.

%Filtered thermal. P= \hbar\omega_0 n \ell. 
%\dot{S} = \kB \ell \log(2n) = \kB \ell \log (2P / \hbar\omega_0 \ell). ????

\section{Summary and conclusions}
%%\label{}
Calculating the entropy flow for a laser beam fills a gap in knowledge. The result is remarkably simple, and has an intuitive explanation. 
It gives new insight into the nature of laser light and confirms, in a new form, its important differences from (unidirectional) thermal light~\cite{Wis16a}. 

Because laser light is far from thermal equilibrium, the entropy it carries has a statistical, rather than directly thermodynamic, interpretation. However, Landauer’s principle~\cite{Landauer1961,Bennett2003} implies that there would be a thermodynamic power cost,  proportional to the entropy flow rate we derive, to erase the information in the beam. This could be interpreted as a lower bound on the power cost to  phase-stabilize a laser by monitoring its output relative to a phase reference.

Another direction for future work would be the analogy between an oscillator (such as a pendulum) for mechanical clocks and a laser mode for optical clocks~\cite{PhysRevFocus:12.114}. In this context,  the result here may also shed light on the question of irreversibility in clocks~\cite{Milburn_2020,Oxford2025}.

\section*{Acknowledgment}
The author thanks Lucas Ostrowski and Kiarn Laverick for valuable comments, and acknowledges the use of M365 Copilot to help solve maths problems and find relevant literature. Any results not available from standard mathematical tables of integrals have been checked by the author using theorems or results in the references cited. This work was supported by Australian Research Council Discovery Project DP220101602.

%After proofreading the manuscript, compress your .tex manuscript file and all figures (which should be in EPS or PDF format) in a ZIP, TAR, or TAR-GZIP package. All files must be referenced at the root level (e.g., file \texttt{figure-1.eps}, not \texttt{/myfigs/figure-1.eps}). If there are supplementary materials, the associated files should not be included in your manuscript archive but be uploaded separately through the Prism interface.

%%%%%%%%%%%%%%%%%%%%%%% References %%%%%%%%%%%%%%%%%%%%%%%%%

%%%%%%%%%% If using BibTeX:
\bibliographystyle{apsrev4-2}
\bibliography{example}

% Full bibliography added automatically for Optics Letters submissions; the following line will simply be ignored if submitting to other journals.
% Note that this extra page will not count against page length
%\bibliographyfullrefs{example}

\end{document}